\documentclass[%
 aip,
 amsmath,amssymb,
 reprint,%
]{revtex4-1}

\usepackage{graphicx}
\usepackage{dcolumn}
\usepackage{bm}

\usepackage[utf8]{inputenc}
\usepackage[T1]{fontenc}
\usepackage{mathptmx}
\usepackage{etoolbox}

\makeatletter
\def\@email#1#2{%
 \endgroup
 \patchcmd{\titleblock@produce}
  {\frontmatter@RRAPformat}
  {\frontmatter@RRAPformat{\produce@RRAP{*#1\href{mailto:#2}{#2}}}\frontmatter@RRAPformat}
  {}{}
}%
\makeatother

\begin{document}

\title{Polarized tip-enhanced Raman spectroscopy at liquid He temperature in ultrahigh vacuum using an off-axis parabolic mirror}

\author{L. Peis}
\affiliation{Walther Meissner Institut, Bayerische Akademie der Wissenschaften, 85748 Garching, Germany}
\affiliation{School of Natural Sciences, Technische Universit\"at M\"unchen, 85748 Garching, Germany}
\affiliation{IFW Dresden, Helmholtzstrasse 20, 01069 Dresden, Germany}

\author{G. He}
\email{gehe1216@gmail.com}
\altaffiliation{Present address: Physics department, University College Cork, College Road, Cork T12 K8AF, Ireland}

\affiliation{Walther Meissner Institut, Bayerische Akademie der Wissenschaften, 85748 Garching, Germany}

\author{D. Jost}
\altaffiliation{Present address: Stanford Institute for Materials and Energy Sciences, SLAC National Accelerator Laboratory, 2575 Sand Hill Road, Menlo Park, CA 94025, USA}
\affiliation{Walther Meissner Institut, Bayerische Akademie der Wissenschaften, 85748 Garching, Germany}
\affiliation{School of Natural Sciences, Technische Universit\"at M\"unchen, 85748 Garching, Germany}


\author{G. Rager}
\affiliation{Walther Meissner Institut, Bayerische Akademie der Wissenschaften, 85748 Garching, Germany}
\affiliation{School of Natural Sciences, Technische Universit\"at M\"unchen, 85748 Garching, Germany}

\author{R. Hackl}
\email{hackl@tum.de}
\affiliation{Walther Meissner Institut, Bayerische Akademie der Wissenschaften, 85748 Garching, Germany}
\affiliation{School of Natural Sciences, Technische Universit\"at M\"unchen, 85748 Garching, Germany}
\affiliation{IFW Dresden, Helmholtzstrasse 20, 01069 Dresden, Germany}

\date{\today}

\begin{abstract}
Tip-enhanced Raman spectroscopy (TERS) combines inelastic light scattering well below the diffraction limit down to the nanometer range and scanning probe microscopy and, possibly, spectroscopy. In this way, topographic and spectroscopic as well as single- and two-particle information may simultaneously be collected. While single molecules can now be studied successfully, bulk solids are still not meaningfully accessible. It is the purpose of the work presented here to outline approaches toward this objective. We describe a home-built, liquid helium cooled, ultrahigh vacuum tip-enhanced Raman spectroscopy system (LHe-UHV-TERS). The setup is based on a scanning tunneling microscope and, as an innovation, an off-axis parabolic mirror having a high numerical aperture of approximately $0.85$ and a large working distance. The system is equipped with a fast load-lock chamber, a chamber for the  \textit{in situ} preparation of tips, substrates, and samples, and a TERS chamber. Base pressure and temperature in the TERS chamber were approximately $3\times 10^{-11}$~mbar and 15~K, respectively. Polarization dependent tip-enhanced Raman spectra of the vibration modes of carbon nanotubes were successfully acquired at cryogenic temperature. Enhancement factors in the range of $10^7$ were observed. The new features described here including very low pressure and temperature and the external access to the light polarizations, thus the selection rules, may pave the way towards the investigation of bulk and surface materials.

\end{abstract}

\maketitle

\section{Introduction}

Tip-enhanced Raman spectroscopy (TERS) integrates scanning probe microscopy with Raman spectroscopy, consequently enabling the acquisition of nanoscale inelastic photon scattering fingerprints of matter \cite{Pettinger:2002, Lucas:2012, Pozzi:2017, Jeremy:2020}. The original concept was proposed by John Wessel in 1985, where a surface plasmon resonance in a metallic tip having a typical apex radius in the range of a few nm. The geometry yields a strong spatial confinement of the electric field at the tip apex and enables the use of evanescent waves. This allows studies of a nearby sample surface in volumes much smaller than given by the diffraction limit \cite{Wessel:1985}.

It took another 15 years before the first TERS experiments were reported independently by four groups \cite{Raoul:2000,Bruno:2000,Norihiko:2000,Anderson:2000} triggering the rapid development of TERS techniques. On the one hand, by optimizing design, tips and substrates, the TERS enhancement factor was improved significantly \cite{Zhang:2013,Wickramasinghe:2014,Chiang:2016,Sheng:2018} towards the theoretical limit of approximately $10^{12}\,$ \cite{Bruno:2010}. On the other hand, the spatial resolution reached nanometer or even $\text{\AA}$
ranges through improvements in the stability of the tip-sample configurations \cite{Zhang:2013,Zhang:2019}. Until now, the vast majority of TERS studies focused on molecules either in ambient conditions or under ultrahigh vacuum (UHV) and cryogenic temperatures \cite{Pozzi:2017, Jeremy:2020}. By taking advantage of increased enhancement factors due to the so-called "gap mode" between a tip and a metal surface, e.g. single crystal Au or Ag, in addition to substantially reducing the far-field using optimized optics such as on-axis parabolic mirrors and radially polarized light \cite{Steidtner:2007,Sackrow2008}.

The introduction of TERS to the study of bulk and thin-film materials potentially facilitates the simultaneous observation of single-particle tunneling spectra and two-particle Raman spectra in the same field of view with atomic resolution \cite{Liu:2019}.  However, several challenges need to be fixed: First, due to the excitation of the complete illuminated bulk volume, the contrast between near and far field is decreased by a factor of approximately $10^{-9}$, as for example shown for Si \cite{Steidtner:2007,Sheng2017} and MoS$_2$ \cite{Voronine:2017}. Second, the requirements of UHV and low temperature for investigating quantum states, such as superconductivity, magnetism, and novel topological states, entail increasingly complicated TERS setups. So far, only few TERS instruments have been demonstrated to enable experiments in UHV conditions and at either liquid nitrogen \cite{Sheng:2018} or liquid helium \cite{Klingsporn:2014} temperatures. TERS facilities achieving sufficiently high enhancement factors and providing access to the Raman selection rules as well as being compatible with low temperatures and UHV are still at their infancy.


In this publication, we describe a home-built ultrahigh vacuum tip-enhanced Raman scattering system having a liquid-He-cooled sample stage (LHe-UHV-TERS), a scanning tunneling microscope (STM) and an off-axis parabolic mirror. In carbon nanotubes, we observed  Raman signals with gain factors $g\approx10^7$ (tip enhancement factors normalized to the scattering volume) at 15\,K and 3$\times$10$^{-11}$\,mbar. In addition, we studied the access to the Raman selection rules. To our knowledge, this is the first demonstration of a parabolic-mirror-based TERS at liquid helium temperature and UHV conditions. Given the high numerical aperture, the access to clean polarization configurations, and the cryogenic compatibility of a parabolic mirror \cite{Steidtner:2007,Nitin:2017,Pena:2022}, our work paves the way to the application of TERS in bulk materials.

\begin{figure}[ht!]
	\centering
	\includegraphics[width=8.5 cm]{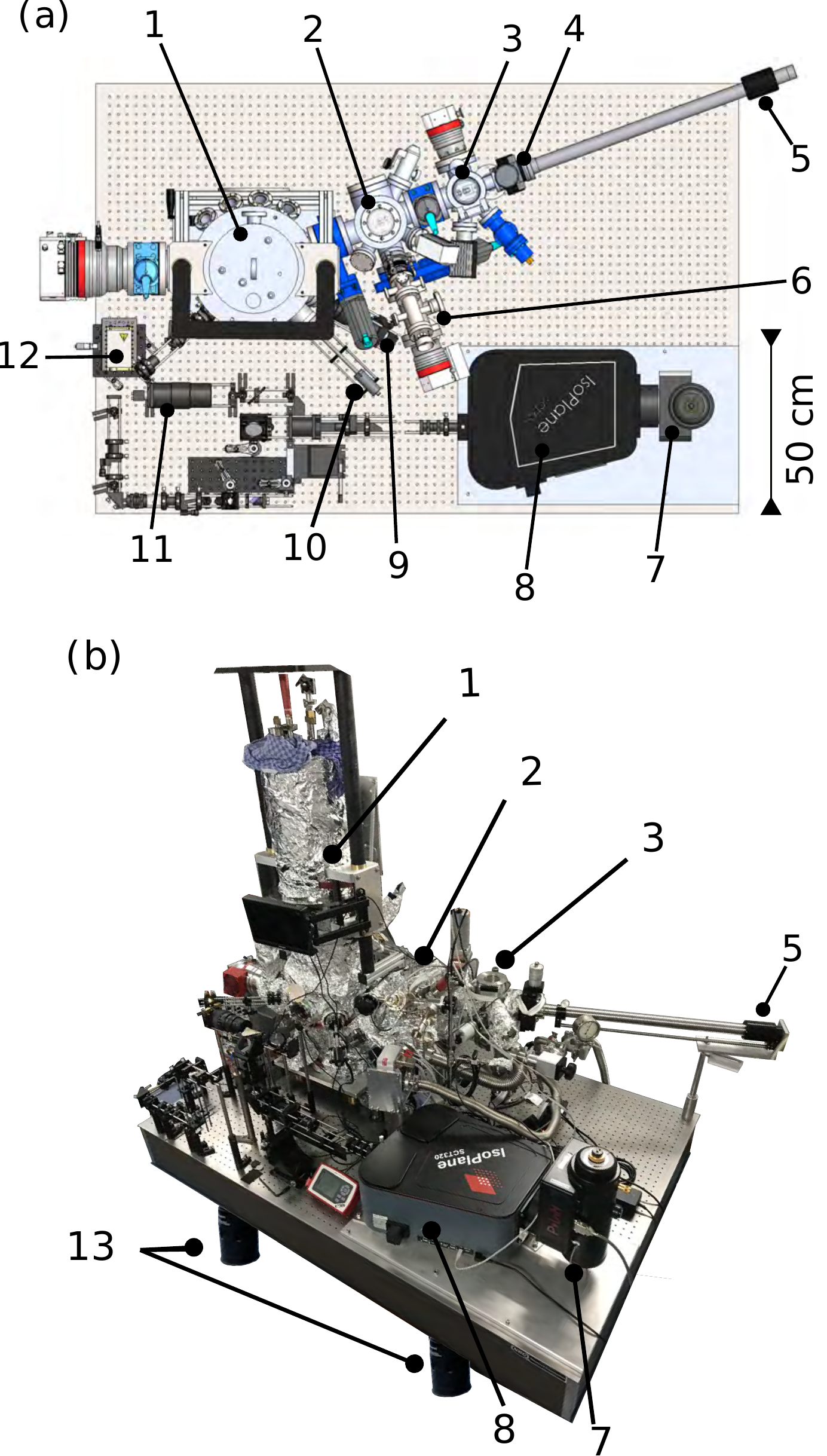}
	\caption{ \textbf{Blueprint (a) and photograph (b) of the customized LHe-UHV-TERS.} The setup consists of the TERS chamber combined with cryogenic equipment (1), a sample preparation chamber (2), enabling annealing and cleaning via a home-made heater and ion sputtering (6), respectively, and a small load-lock chamber (3) for short initial pumping times. The transfer is facilitated utilizing a height-adjustable (4) transfer arm (5) and a wobble stick (10). Some other key components are: (7) CCD, (8) spectrometer, (12) laser mount, (13) pneumatic vibration isolation system, and (9, 11) two observation cameras.}
	\label{figTERS3Dmodel}
\end{figure}


\section{General design}

The objective was an experimental setup for polarization-resolved Raman scattering in far- and near-field configuration working at helium temperature and UHV conditions. For providing enough space for the scanning stage and the \textit{in situ} exchange of samples and silver tips we introduced an off-axis parabolic mirror.

A top-view blueprint and a photograph of the system are shown in Fig. \ref{figTERS3Dmodel}. The whole setup is installed on an optical table supported by four vibration isolators (Newport, S-$2000\,$ stabilizer). The system consists of three chambers having different functions: (1) TERS chamber, (2) preparation, and (3) load lock (see Fig. \ref{figTERS3Dmodel}(a) for details). The load-lock chamber is the access point for samples and tips. The subsequent preparation chamber equipped with an ion sputtering gun and a homemade annealing stage (up to approximately 600 degrees) allows the $in$-$situ$ cleaning of samples and tips. Experiments are carried out in the TERS chamber which integrates a multi-axis STM head (Attocube), an off-axis parabolic mirror, and an UHV cryostat (Janis Research, CNDT series). The three chambers are individually pumped and separated by UHV valves.

Ultrahigh vacuum (UHV) is achieved by a serial pumping system. The TERS chamber is pumped by a 300\,liter turbo pump (HiPace 300, Pfeiffer Vacuum), which itself is pre-pumped by a dry Scroll pump (HiScroll 12, Pfeiffer Vacuum). If needed, an additional ion getter pump can be installed on a CF150 flange. Both the preparation chamber and load-lock chamber are equipped with 80\,liter turbo pumps (HiPace 80, Pfeiffer Vacuum) with additional dry fore pumps.

After baking the system at 150 degrees ~for three days base pressures are obtained reaching $8\times10^{-9}\,$mbar in the TERS chamber, $3\times10^{-8}\,$mbar in the preparation chamber and $5\times10^{-8}\,$mbar in the load-lock chamber. After cooling down to He temperature, a final minimal pressure of $1.4\times10^{-11}\,$mbar could be achieved in the TERS chamber.

The UHV cryostat is filled with 7\,L liquid nitrogen in the outer tank and 4\,L liquid helium in the inner tank. While the sample stage may be operated at temperatures from 15 to 320\,K, it was kept at approximately 20\,K for most of the experiments. To reduce the vibrations induced by the bubbling of liquid nitrogen, the nitrogen tank was pumped down to around 10\,mbar with an additional scroll pump. At 10~mbar the nitrogen solidifies and the noise from the bubbles disappears. The LN$_2$ had to be pumped for the entire measurement runs.

\begin{figure*}[ht!]
	\centering
	\includegraphics[width=13cm]{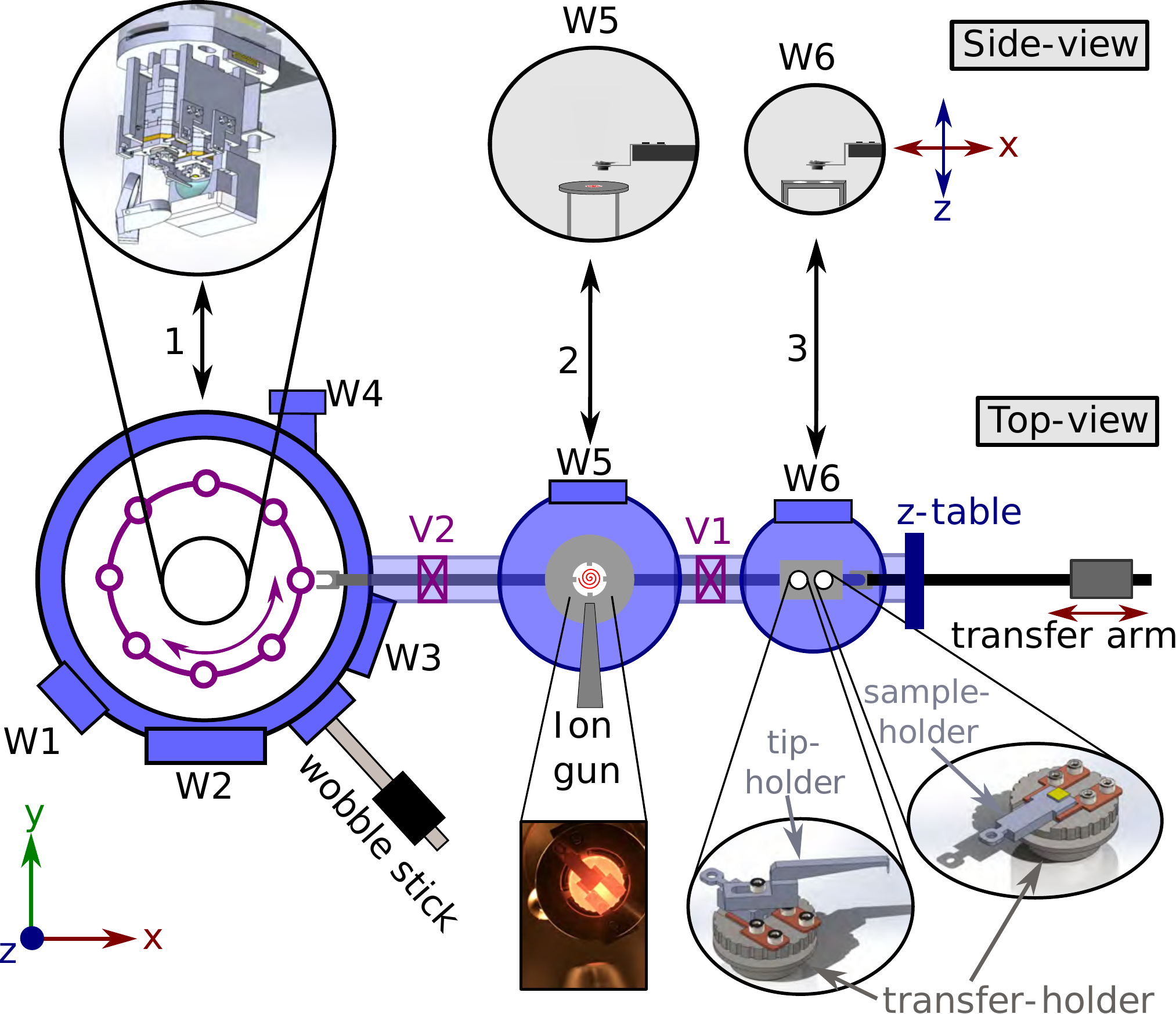}
	\caption{ \textbf{Schematic view of the setup.} The upper row shows a perspective view of the STM part with the parabolic mirror on the left (encircled) and a side view along the $y$-axis as indicated by the inset on the right. The central row displays the top view along the $z$-axis (see axes on the lower left). Left: TERS chamber, middle: preparation chamber, right: load lock chamber and transfer arm. The magnetically coupled transfer arm connects all three chambers and facilitates the loading and transfer of samples and tips. The height of the arm can be adjusted via the $z$-axis manipulator. The samples and tips are stored on a pre-cooled, rotatable carousel with eight positions in the TERS chamber (purple). From there they are being picked up by the wobble-stick and inserted into the STM head. Two valves (V1, V2) separate the individual chambers. The windows W(2-6) enable visual control in addition to W1 and W2 being used for sample illumination and Raman measurements, respectively.}
	\label{figtransfer}
\end{figure*}

\section{Chambers and transfer system}

Figure \ref{figtransfer} shows a schematic view of the three relevant chambers, TERS (left), preparation(middle), and load lock (right). A transfer arm and a wobble stick have been attached to the load-lock and TERS chamber, respectively, enabling the smooth $in$-$situ$ transfer between the individual chambers.

The sample and tip are mounted in flag-style holders (sample and tip holder as indicated in Fig. \ref{figtransfer}), which can be grabbed by a commercial wobble stick. Then, the sample or the tip holder is inserted into a cylindric holder (transfer holder as shown in Fig. \ref{figtransfer}), with the latter being designed to be forked by the transfer arm. Two transfer holders may then be put into the load-lock chamber and inserted into the available two voids of the storage table. A third holder may stay in the fork. So in total three holders may be pumped simultaneously. The transfer arm has a range of 1.5\,m along $x$-direction and can be fine adjusted in $z$ direction by a shaft guidance (4), having a range of 25\,mm. A fork is installed at the end of the transfer arm, in order to pick up the transfer holder and deliver it to the preparation or TERS chamber.

In the preparation chamber, a homemade annealing stage heated by tungsten coils is mounted at the bottom. By applying a current (maximum 5.6\,A) through the coils, the entire transfer holder heats up and starts glowing as seen in the inset of Fig. \ref{figtransfer}. The annealing temperature can reach approximately 600 degrees.

Additionally, an ion gun is installed in the preparation chamber having a tilt angle of $45^{\circ}$ with respect to the sample plane. The ion gun is differentially pumped and operated with $99.995\%$ pure Ar at a chamber pressure of approximately $5\times10^{-5}\,$mbar, controlled via a needle valve. Atomically flat surfaces of gold and silver crystals were achieved after several cycles of alternating ion sputtering and annealing. The ion sputtering approach can also be used for a first cleaning procedure of the tip.

Samples and tips are finally transferred into the TERS chamber. A storage carousel with eight positions is mounted on the bottom of the liquid nitrogen radiation shield and can be turned by the wobble stick. For tip and sample transfer to the STM head, the carousel has to be locked to the radiation shield by screws. From the arrested carousel, the wobble stick grabs the sample/tip holder and inserts it into the sample stage or tip stage on the STM head.

The storage carousel substantially improves the space and time efficiency and also simplifies the design of the STM chamber compared to other designs \cite{Wu:2018}.

\begin{figure*}[ht!]
	\centering
	\includegraphics[width=16cm]{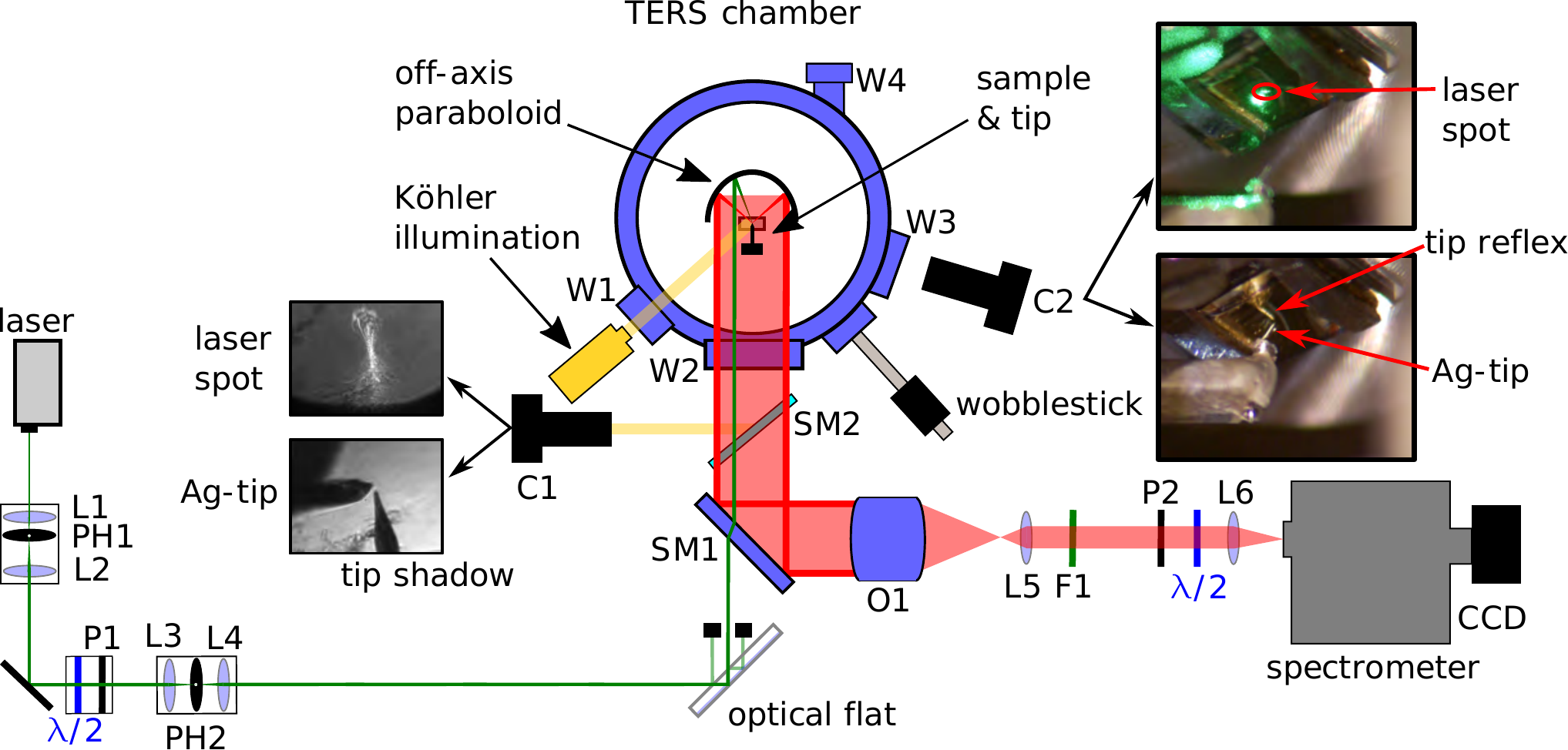}
	\caption{ \textbf{Schematic of the optical configuration for Raman scattering.} The parabolic mirror, sample, and tip are located in the center of the TERS cryostat. The laser for excitation on the left is linearly polarized. Two confocal lenses (L1 and L2) expand the beam to 6~mm. A pinhole (PH1) in their joint focus filters the light spatially. After the mirror, the desired polarization and power are set (P1 and $\lambda/2$). Before hitting the optical flat the beam is again spatially filtered (L3, PH2, L4). The optical flat is adjustable in a way that the beam can be aligned parallel and at the required distance to the optical axis of the parabolic mirror. The paraboloid focuses the light on the sample and collects the inelastically scattered, red-shifted light. The parallel Raman light is deflected towards lens O1 by a semitransparent mirror (SM1). The focal length of lens L5 is smaller by approximately a factor of three in comparison that of O1 thus leading to a reduction of the beam diameter to 12~mm. Before the Raman light enters analyzer P2 the laser is cut off by an edge filter (F1). Another $\lambda/2$ retarder rotates the selected polarization into the direction optimal for the spectrometer. Lens L6 focuses the light on the entrance slit. The K\"ohler illumination system and parts of the observation optics are also illustrated. Tip, sample and laser spot may be imaged by two independent systems (C1, C2). System C1 looks into the paraboloid through the semitransparent mirror SM2 and uses part the paraboloid as an objective lens. Due to the off-axis conditions the aberrations are substantial, and the laser spot is elongated (upper left inset). When the tip gets close to the sample surface the tip and its shadow approach each other (lower left inset). System C2 provides a direct image of the sample with adjustable magnification (upper right inset with laser, lower right inset without laser).}
	\label{figopticalpath}
\end{figure*}

\section{Optical path}

A schematic drawing of the optical setup is displayed in Fig. \ref{figopticalpath}. A diode-pumped solid-state laser emitting at $532\,$nm (Coherent sapphire SF $532$) was utilized for all measurements. The mounting platform allows an easy exchange of the green sapphire laser to a red laser emitting at $660\,$nm (Laser Quantum, Ignis-FS $660$) if desired. The laser beam is spatially filtered and expanded to a diameter of $6\,$mm by the first pinhole system (L1, $f=10$~mm; PH1, $d = 30$~$\mu$m; L2, $f=60$~mm), reducing the divergence of the beam to $2\times10^{-4}\,$rad. Subsequently, the laser power and the polarization direction of the light are selected by the combination of a $\lambda/2$ waveplate and a polarizer (P1).
After the polarizer, the beam is spatially filtered again by a pinhole system (L3, $f=50$~mm; PH2, $d = 30$~$\mu$m; L4, $f=60$~mm) before being directed to an adjustable optical flat (B. Halle Nachfl., Berlin). The latter is a fused silica cuboid, polished to a flatness of $\lambda/20$ and coated with silver on the back surface. Thus the first and the third beam have an intensity of approximately 4\,\% of the incoming laser beam, and the second one has 90\,\%  and serves for excitation of the Raman spectra.  The wedge between the front and back side of the optical flat does not exceed $\lambda/20$ over the entire length of 40~mm, and the three beams are parallel to within $6\times 10^{-7}$~rad. With the central beam blocked, the weaker side beams enable the accurate alignment of the parabolic mirror inside the cryostat, as described in Ref.~\cite{Chen2012}. Cooling-induced changes in the position of the parabolic mirror inside the cryostat can be tackled either by varying the $z$ position of the optical flat or of the cryostat thus maintaining the parallel alignment of the incoming beam and parabola axis.

The last optical element in front of the sample is the parabolic mirror having a solid angle of $\pi$ and a numerical aperture of approximately 0.85. Its diameter and focal length are $30$ and $8\,$mm, respectively. This configuration guarantees a small laser focus. As an innovation, the sample surface is not perpendicular but parallel to the optical axis of the parabola. In this way, the directly back-reflected laser light never enters the collection optics. Both incoming and scattered photons are reflected from the aluminum coating of the  parabola and thus the respective polarizations have undesired components. This  contamination was shown experimentally and theoretically to not exceed 15\,\%. Consequently, the selection rules are by and large effective.
From the focal point, inelastically scattered light is emitted in all directions. The part propagating into the direction of the paraboloid is being collected and re-directed in an ideal parallel fashion. The diameter of 30~mm (corresponding to the diameter of the paraboloid) of the parallel light is reduced to 12~mm by an objective lens (O1, Canon, $f=85\,$mm) and an achromat (L5, $f=40$~mm). Contributions of the elastically scattered light (Rayleigh line) are blocked by an edge filter (F1, Iridian 532nm LPF). The desired polarization state of the inelastically scattered photons is selected by analyzer P2. A $\lambda/2$ wave plate rotates the selected polarization into the direction of the highest sensitivity of the spectrometer.
Finally, the inelastically scattered light is focused (L6, $f=50$~mm) onto the entrance slit of the spectrometer (ISOplane SCT320, Princeton Instruments). The photons transmitted with the desired energy are detected by a charge-coupled device (CCD, PyLoN-400BRX, Princeton Instruments). The spectral range in the green per single exposure for the chip width of 27~mm is approximately 600~cm$^{-1}$. \\
\begin{figure}[ht!]
	\centering
	\includegraphics[width=7cm]{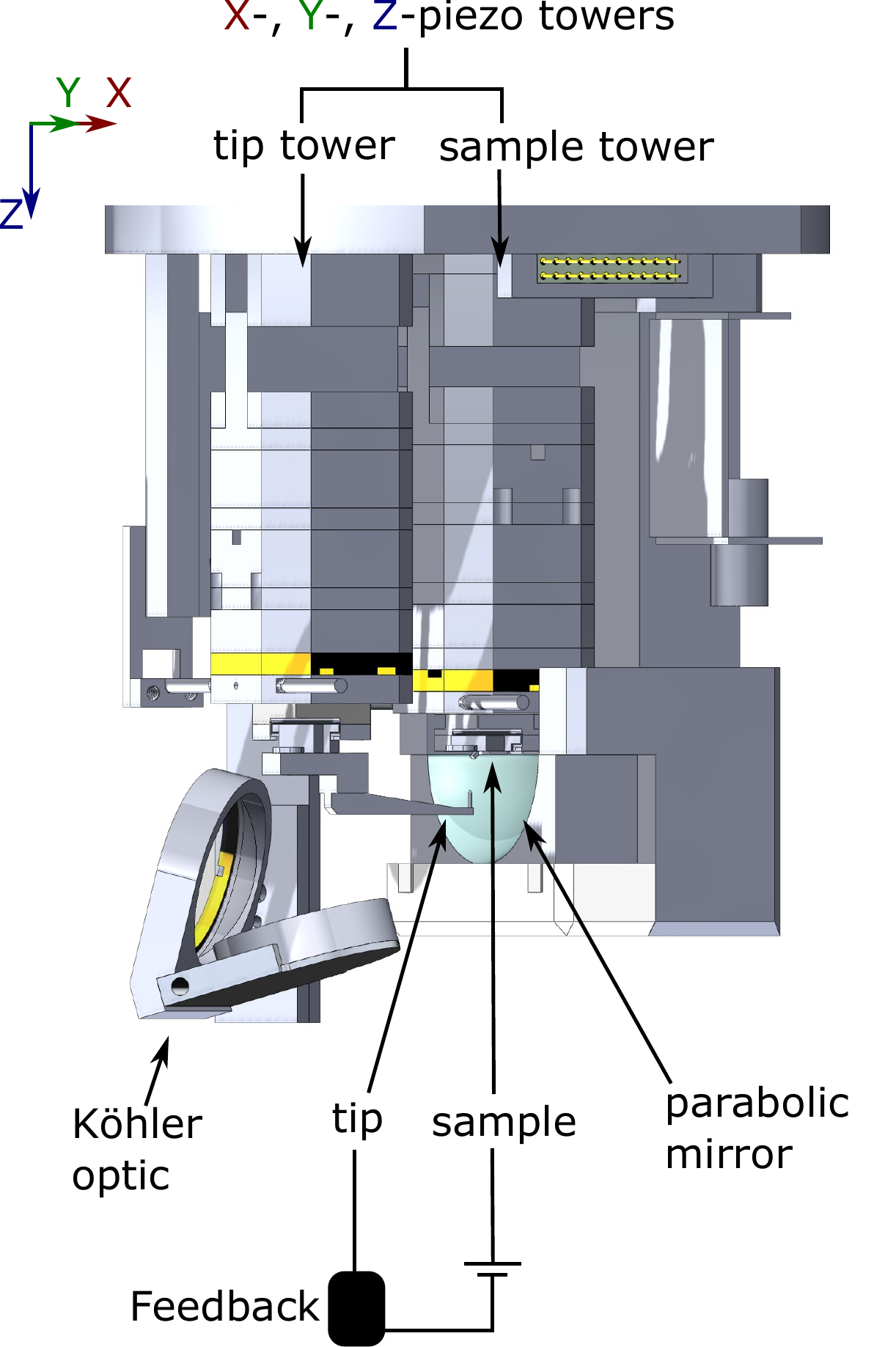}
	\caption{ \textbf{Schematic of the STM head.} The drawing is based on that shown in the upper left part of Fig. \ref{figtransfer}. The front structure has been removed for demonstration purposes. All components (paraboloid, sample tower, tip tower) are mounted on the same platform. All structures are made of titanium. The platform is connected to the He bath by a thermal resistor facilitating the temperature to be varied in the range of 10 to 300~K. Each tower has three linear degrees of freedom.  One of the two operation modes allows sample and tip to be positioned in the focal point of the parabolic mirror. The other operation mode facilitates fine scanning for the measurements. Via a feedback loop, the piezo stages enable the acquisition of STM topographic images. On the lower left, the last lens of the K\"ohler illumination system is displayed along with a mirror that directs the light towards the sample.}
	\label{figSTMschematic}
\end{figure}


The related STM setup is shown in Fig. \ref{figSTMschematic}. The paraboloid on the right hand side is fixed. The sample  may be moved by the sample tower along three independent axes in either coarse mode in a range of $\pm 2.5$~mm or fine scanning  mode in a range of approximately $\pm 2~\mu$m. The tip is mounted on an identical second tower (tip tower). First, the sample is moved into the focus of the paraboloid using the two weak parallel laser beams for fine-adjustment as mentioned above and described in detail in Ref. \cite{Nitin:2017}. Then the tip is approached and moved into the focus.

To this end, the sample and tip region are observed by two independent camera systems (C1, C2), enabling the controlled movement of  sample and tip, the tip approach, and the optical alignment. A large field of view is accessible through window W3 by camera C2. The attached zoom lens (MVL700, ThorLabs) allows the variation of the field of view between $6\times 10$~mm$^2$ and $36\times 60$~mm$^2$. The sample region is illuminated by a K\"ohler system for optimal contrast.

Typical images are shown in the two insets of Fig. \ref{figopticalpath} on the right. The upper inset shows the gold-plated silicon sample with the laser spot (red circle; the big bright area is a reflex from the window). The arm with the tip is retracted and thus out of focus (lower left part of the inset).  The lower inset shows the sample with the tip approached. The silver tip looks golden because of the sample surface. The reflex of the tip is nearly perpendicular to the tip close to the upper edge of the sample.

The final approach of the tip to the sample is optically controlled by observation system C1. The zoom objective (Sigma Optics) is set at infinity and uses a semi-transparent mirror (SM2) for looking into the paraboloid which is used as an objective lens. In this way  microscopic observation of sample surface, tip, and laser spot with $\mu$m-resolution becomes possible. Resulting images are depicted in the left insets in Fig. \ref{figopticalpath}. The tip and its shadow can be seen clearly (lower left inset) allowing a coarse approach down to approximately  $1~\mu$m. The laser spot (upper left inset) is elongated because of the geometrical aberrations induced by the off-axis conditions. Its lateral size is close to $5~\mu$m. All dimensions may only be estimated relative to the tip thickness.

\begin{figure*}[ht!]
	\centering
	\includegraphics[width=14cm]{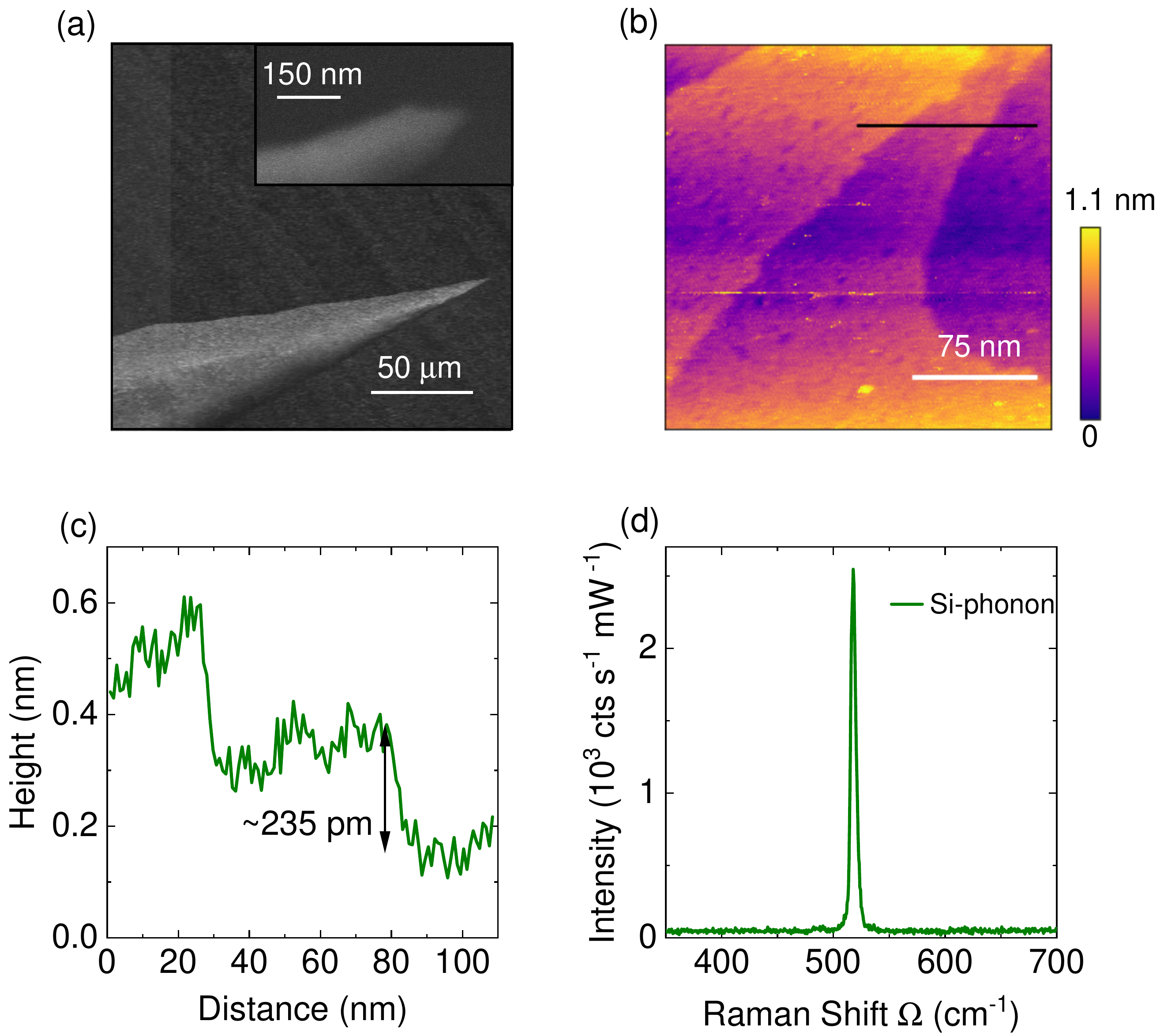}
	\caption{ \textbf{Raman and STM performance of the LHe-UHV-TERS system.} (a) Electrochemically etched Ag-tip with a tip apex radius of around $50\,$nm. (b) STM image of a gold surface where the height is color coded. The black line indicated the position of the line cut in panel (c). (c)  The terraces are flat to within the noise level of approximately 0.5~\AA. The steps between the terraces can be clearly resolved. (d) Raman spectra of a (001)-Si surface in $ab$-polarization configuration. For the Si phonon at $520\,$cm$^{-1}$ the count rate in the peak is approximately 2500~cts/(s mW).}
	\label{figperformance}
\end{figure*}


\section{Preparation of the tip}
Since Ag tips show the strongest surface plasmon excitation for green wavelengths while having negligible luminescence (as opposed to Au), we switched to Ag. Yet, Ag is much more tedious to handle because of the instability of the surface. We first prepared the tips using electrochemical etching as described in Ref. \cite{Nitin:2017}. Figure \ref{figperformance}(a) displays the SEM image of a typical Ag-tip having a radius of less than 20~nm. One of those tips was selected and tested before being transferred into the STM stage. The preparation is among the crucial parts of the experiment. The tip  is conditioned in close vicinity of either a Au single crystal or pristine thin film using field emission. For this purpose, we apply a voltage between the substrate and tip of up to 75\,V so that the current reaches 10\,$\mu$A for up to 30 minutes. The tip can be further improved by short voltage pulses. After that, the tip quality and cleanliness can be checked by STM.

\section{Performance of STM and Raman}
The performance of STM is illustrated by the topographic image of a single crystalline Au(111) film shown in Fig. \ref{figperformance}(b). The atomic terraces and the steps are readily resolved as shown in the line cut in Fig. \ref{figperformance}(c). The height of a mono-layer atomic step (235\,pm \cite{Walen2015}) is used to calibrate our STM head. This measurement and those shown below were performed with a tunneling current of 1~nA and a bias voltage of 500~mV. Further checks are possible upon scanning the topography of SWNTs on the gold substrate.

\begin{figure*}[ht]
	\includegraphics[width=16cm]{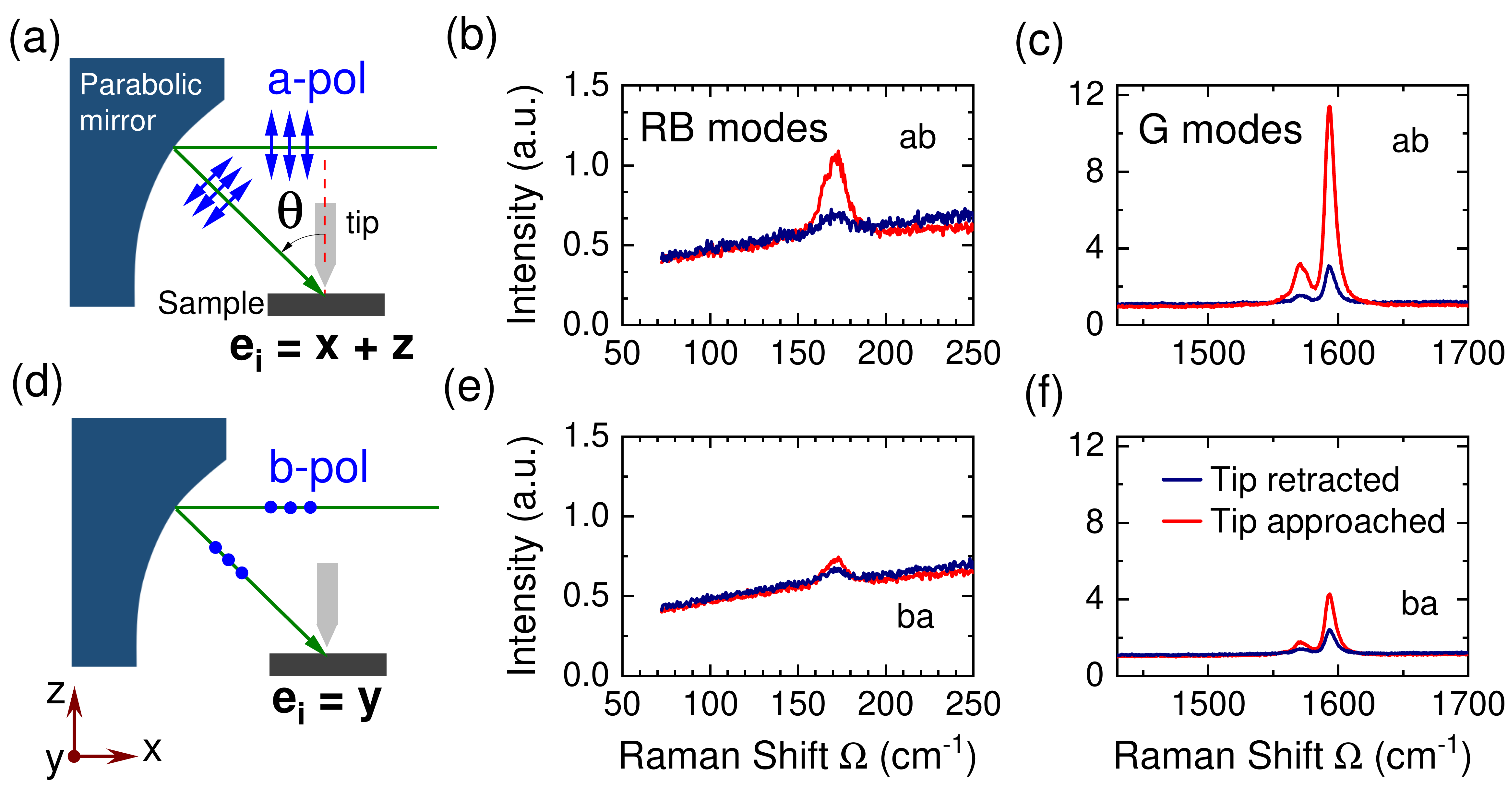}
	\caption{ \textbf{Polarization-dependent enhancement of the Raman signal.} (a) and (d): Definition of the polarization configurations. Vertical and horizontal polarizations parallel to the laboratory $z$ and $y$ axes are labeled $a$ ($a$-pol) and $b$ ($b$-pol), respectively. The incoming polarization ${\bf e}_i$ of the photons on the sample surface depends on both the polarization of the incoming light ($a$ or $b$) and the angle of the beam with respect to the tip axis $\theta$. The enhancements $EF$ for $ab$ and $ba$ configuration are depicted in the upper row and, respectively, lower row for the radial breathing modes [(b) and (e)] and the G modes [(c) and (f)]. For better visualization, the background signals were normalized for the two configurations. All modes are enhanced in approximately the same way when the polarization axis is aligned along the tip. This dichotomy may allow one to distinguish contributions from far-field and near-field, as well as $c$-axis contributions in bulk and thin film materials in future studies.}
	\label{figpolarizationTERS}
\end{figure*}

Before combining STM and Raman spectroscopy we measured the intensity of the Si phonon at 520~cm$^{-1}$ in the TERS system and compared it with earlier results obtained in a scanning spectrometer. The experiments were carried out on $\{001\}$ surfaces. An $ab$ polarization configuration (for the definition see Fig.~\ref{figpolarizationTERS}) was chosen for measurements covering a range of 338 to 1028~cm$^{-1}$. The point density is given by the pixel size of the CCD. For the wavelength selected, $20~\mu$m corresponds to approximately 0.5~cm$^{-1}$. The laser power absorbed by the sample and the dwell time for a single acquisition were set at 1~mW and 1~s, respectively. The resulting spectrum is plotted in Fig. \ref{figperformance}(d). The peak intensity reaches $\approx2500~$cts/(s mW). The corresponding peak intensity in the scanning spectrometer using a resolution of 5~cm$^{-1}$ is 7000~cts/(s mW). With all optical elements and the different resolutions in the two experiments properly taken into account, the sensitivity of the TERS setup is at least a factor of two higher owing to the higher aperture of the parabola.  

\begin{figure*}[ht!]
	\centering
	\includegraphics[width=14cm]{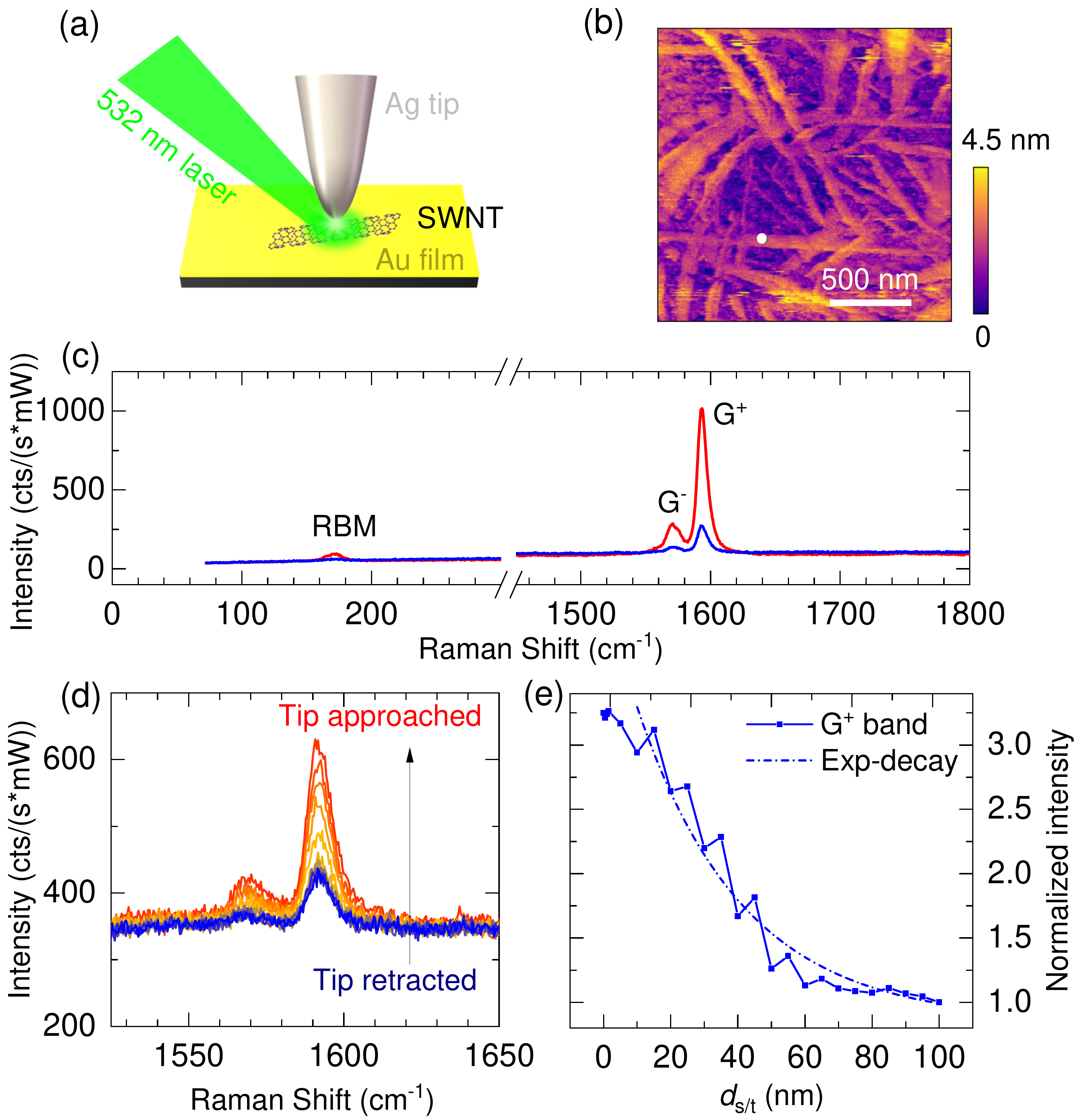}
	\caption{ \textbf{Tip enhanced Raman signal of SWNTs at $20\,$K.} (a) Close-up of the general setup of the sample, tip, and laser. (b) Topographic image of the SWNTs sample, illustrating several different nanotube bundles on the gold surface. The tip was positioned on an isolated region of an individual SWNT bundle (white dot). (c) Enhancement of several characteristic vibration modes of SWNTs. (d) Raman spectra visualizing the distance dependence of the TERS signal. From the normalized intensity of the G$^+$ excitation, the exponential decrease upon removing the tip can be observed, as depicted in panel (e).}
	\label{figTERSG+G-bands}
\end{figure*}
\section{Enhancement in single wall carbon nanotubes}

Polarization-resolved Raman spectra can be measured with very high sensitivity in the temperature range between 10 and 300~K under UHV conditions using a parabolic mirror as shown in Fig. \ref{figperformance}(d). Both Au and Ag tips were prepared routinely and may be transferred into the measurement position. The STM has a $z$ resolution of approximately 50~pm (0.5~\AA) as shown in Fig. \ref{figperformance}(c). These features allow us to perform test measurements using single-wall carbon nanotubes (SWNT) as samples. Although these experiments are not new (see refs. \cite{Liao2016,Chen2014,Peica2011}) they help us to pin down the performance of our system and compare it to earlier experiments.

SWNTs are in fact much easier to deal with than surfaces of crystals \cite{Steidtner:2007} since they intrinsically limit the scattering volume (in a fashion similar to single molecules) and allow the direct visualization of the tip enhancement in the Raman spectra. In addition, the samples are easy to prepare and stable at ambient conditions.

SWNTs are suspended in H$_2$O. A drop of the suspension is dripped on a gold film in ambient air. The liquid is evaporated in a continuous flow of  clean nitrogen for two minutes. The successful deposition of the SWNTs is validated using an atomic force microscope (AFM) under ambient conditions. Finally, the gold film with the attached SWNTs is loaded into the chamber.

A silver tip and, as above, a green laser emitting at 532\,nm are used. Figure \ref{figTERSG+G-bands}(a) displays the configuration of our TERS measurements.  An STM image of the sample with the SWNTs is shown in Fig. \ref{figTERSG+G-bands}(b). It is difficult to resolve individual nanotubes within our resolution limit, but rather several bundles with each containing several nanotubes \cite{Hartschuh:2003}.

From this starting point, we move the tip to one of the bundles as indicated by the white spot in Fig. \ref{figTERSG+G-bands}(b), and measure Raman spectra (see Fig. \ref{figTERSG+G-bands}(c)). In the far field with the tip retracted, one can identify the radial breathing mode (RBM) at 170\,cm$^{-1}$, and the G$^-$ and G$^+$ modes at $\Omega \approx 1569.7$ and $1593.8$~cm$^{-1}$, respectively. Once the tip is approached to the sample surface, all three modes are enhanced substantially by factors in the range between 4 and 6 whereas the background changes only negligibly. This stability of the background is another criterion for the quality of the tip. If the Ag tips are not appropriately prepared the background always exhibits a substantial increase upon tip approach. Further cleaning along the lines described above removes this additional luminescence.

For a quantitative analysis of the Raman intensity, the phonon peaks were fitted with Lorentzians and a flat background, giving the ratio of the peak intensities, defined here as the enhancement factor $EF$ :
\begin{equation}
	EF=\frac{I_{\text{approached}}}{I_{\text{retracted}}}.
	\label{eq:TERSEF}
\end{equation}

Note that this is the actual intensity enhancement of the investigated Raman processes. For the determination of the gain $g$ induced by the tip the scattering volumina relevant for the near and far field scattering have to be taken into account \cite{Steidtner:2007}.  The gain $g$ is given by:
\begin{equation}
	g = c*\left(\frac{I_{\text{total}} - I_{\text{far}}}{I_{\text{far}}} \right)*\frac{\text{A}_\text{far}}{\text{A}_\text{near}} ,
	\label{eq:TERSg}
\end{equation}
where $c$ is the coverage of SWNTs and $A_\text{far}/ A_\text{near}$ is the ratio of the contributing areas $A_i$ with the tip retracted (far) and approached (near). We can use the ratio of areas instead of the complete illuminated volumes since the contributing SWNTs are quasi-2D in our measurements. The coverage ratio may be estimated from the STM topographic image as described in more detail in Appendix A. The resulting values for $EF$ and $g$ are listed in Table \ref{tab:Enhancementfactor}. The value for $g$ of order $10^7$ found here for the G$^+$-band of SWNTs is significantly larger than those described in the literature. Most publications report $g$-factors in the range of $10^4$ to $10^6$ \cite{Shiotari2014,Vannier2006,Kato2020,Pettinger2012}. Recently, gains of $10^8$ were observed by Liao et. al \cite{Liao2016}. We conclude that the LHe-UHV-TERS-system described here meets the benchmarks.

\begin{table}[ht]
	\caption{\textbf{Intensity enhancement $EF$ and TERS gain $g$ of SWNT excitations.} $EF$ and $g$ are calculated by applying Eqs. \ref{eq:TERSEF} and \ref{eq:TERSg} to the data presented in Figs. \ref{figTERSG+G-bands} (c) and \ref{figpolarizationTERS} for both $ab$ and $ba$ polarization.}
	\centering
	\setlength{\tabcolsep}{5mm}{
		\begin{tabular}{c c c c}
			\hline
			 Excitation& & $ab$ & $ba$ \\
			 \hline
			 \hline
			{RBM} & $EF$ & $4.88$ & $2.53$ \\
			& $g$ & $1.94\times 10^7$ & $7.65\times 10^6$ \\
            {G$^-$-band} & $EF$ & $5.58$ & $2.53$ \\
			& $g$ & $2.29\times 10^7$ & $7.65\times 10^6$ \\
            {G$^+$-band} & $EF$ & $$5.74$$ & $2.68$ \\
			& $g$ & $2.37\times 10^7$ & $8.40\times 10^6$ \\
			\hline
	\end{tabular}}
	\label{tab:Enhancementfactor}
\end{table}

Next, the Raman spectra are studied as a function of tip/sample distance. Starting from the tunneling mode we retract the tip slowly to $100\,$nm with a step size of $5\,$nm by reducing the $dc$ voltage applied on the $z$-axis scanner of the tip tower. At each step, a Raman spectrum was taken and is displayed in Fig. \ref{figTERSG+G-bands}(d). For close distances of the tip to the sample surface, the intensities of the G$^+$ and G$^-$ modes are increased clearly. The distance-dependent intensity of the statistically more robust G$^+$ mode is plotted in Fig. \ref{figTERSG+G-bands}(e), which can be fitted using an exponentially decaying function. We find the distance for a decrease to 1/e at 32\,nm, close to the values reported in \cite{Kharintsev:2007}.

\section{Polarization dependence}

The Raman selection rules allow the identification of excitations. In addition, far- and near-field responses have to be separated. Thus, access to the light polarizations is crucially important. However, the direction of the tip limits the flexibility since the enhancement is much larger for light polarizations along the tip axis. Nevertheless, for the potential use of the selection rules and the distinction of far and near-field intensities we study the influence of the light polarizations on the intensities of the vibrations of SWNTs.

First, we align the incoming laser beam in a  way that the polarizations can be selectively adjusted to be either parallel or perpendicular to the tip. Since the aluminum coating of the paraboloid generally adds elliptical components to the originally linearly polarized photons special care is necessary to optimize the light path. For best results, the light comes in next to the tip [small azimuthal angle $\varphi$ measured from the plane defined by the tip axis and the optical axis and large polar angle $\theta$ with respect to the tip axis as shown in Fig. \ref{figpolarizationTERS} (a) and (d)] since then the plane of incidence, by and large, includes the tip and parabola axes facilitating the distinction between parallel and perpendicular incident polarizations on the parabola's surface.

Raman measurements were then performed in $ab$ and $ba$ configurations as defined in panels (a) and (d) and the caption of Fig. \ref{figpolarizationTERS}. The resulting spectra for the retracted and tunneling mode are presented in Fig. \ref{figpolarizationTERS} (b), (c), (e), and (f). For better visualizing the enhancement factors, the $ab$ and  $ba$ spectra were divided by factors of $89$ and $228$, respectively. Then, the background of all spectra at low energies is close to 0.5~cts/(s mW) and close to unity for high energies.

The enhancement factors are two times higher in the $ab$ channel than in the $ba$ channel as summarized in Table \ref{tab:Enhancementfactor}. This is consistent with simulation results, where the polarization along the tip axis has the highest evanescent field \cite{Novotny:1997}. This behavior possibly allows the distinction of the far and near field by only changing the polarization outside of the cryostat, at least in a material in which $ab$ and $ba$ map the same Raman symmetries. In such a scenario, the tip is kept at its position, while the processes (enhanced and not) can be studied and compared easily. Thin film samples with fourfold crystal symmetry 
are candidates, e.g.,  monolayer FeSe having an enhanced superconducting transition temperature \cite{Wang:2012}.

An alternative approach for comparing far and near field while maintaining the position of the tip and the  polarization along $a$ is the variation of the polar angle of the incoming light beam $\theta$ with respect to the tip axis. The adjustable height of the optical flat provides comfortable access to $\theta$. Along with the variation of $\theta$ the field enhancement at the tip apex and, consequently, the $c$-axis contribution to the spectra of bulk samples changes. \\

\section{Conclusion}
To summarize, we have described and successfully tested a home-built system for tip-enhanced Raman scattering which operates at temperatures down to 15~K close to the boiling point of liquid helium and in ultra-high vacuum. The cryo-pumped vacuum reached a minimal pressure of $3\times10^{-11}$~mbar. Silver tips were prepared \textit{in situ}. The light is focused on the tip and collected after scattering by an off-axis paraboloid.

For benchmarking purposes, we studied topographic images and Raman spectra of single-wall nanotubes dispersed on a thin gold film. The axial and lateral resolution of the scanning tunneling microscopy is a fraction of an atomic layer (approximately 0.5~\AA) and in the \AA-range, respectively. We demonstrated that the Raman signal of all molecular vibrations of the nanotubes increased exponentially by factors of up to 5.7 upon approaching the Ag tip. We observe that the intensities of all excitations depend also on the photon polarizations. The effect of rotating the incoming polarization from parallel to perpendicular to the tip axis is stronger by a factor of two than that of the rotation of the outgoing polarization. This dichotomy may either result from the polarization dependent ratios of the scattering volumes or the selection rules or from the different antenna effects of the tip for incoming and outgoing photons.

For getting better \textit{in situ} access to the sample and the tip we have changed the orientation of the sample surface from perpendicular to the optical axis of the parabola \cite{Steidtner:2007} to parallel. Since we can use only half of the paraboloid the solid angle is only $\pi$ rather than $2\pi$ although the mounting of the STM and of the sample reduces the free aperture in the latter case. The orientation used here allows reasonable (though not full) access to the light polarizations. The stability of the STM system was not as high as desirable but sufficient for the current proof of principles. Using this equipment we could clearly demonstrate tip enhancement for SWNTs at a temperature much lower than in previous experiments.

The advantage of nanotubes (similar to the case of single molecules) is the confinement of the scattering volume in the far field facilitating the visualization of the enhancement. The big challenge of future work is the application of the technique to bulk samples such as topological insulators having metallic surface states and an insulating bulk. 

\section{Appendix A: Estimation of the enhancement factor $g$}\label{sec:AA}

As opposed to the enhancement factor $EF$ the gain $g$ is not directly accessible. For an estimate, one would need to know the coverage ratio of SWNTs and the related area ratio  $A_\text{far}/ A_\text{near}$ for far and near field. We start with the minimal laser spot diameter which can be calculated as

\begin{equation}
	d = 1.27 \cdot \lambda \cdot M^2 \cdot \frac{f}{D} \approx 900\,\text{nm}
\end{equation}

where $M^2$ is assumed to be $1$ for a perfect Gaussian spot shape, $f$ is 8~mm, the beam diameter $D$ is 6~mm, and $\lambda=532$~nm. While the theoretical limit $d$ is very hard to achieve we estimate the spot size \textit{via} the image taken from camera C1, yielding a more realistic spot diameter of approximately $d^\prime=5~\mu$m. The coverage $c$ of the gold film by SWNT may be estimated on the basis of the STM topographic image in Fig. \ref{figTERSG+G-bands}(b), revealing approximately $c=0.8$. The near field signal confined to the vicinity of the tip is characterized by an area of $\pi r^2$ with $r=1$~nm. Therefore, we can write for the relevant area ratio:
\begin{equation}
     \frac{c \cdot\text{A}_\text{far}}{\text{A}_\text{near}} =  \frac{c \cdot (d^\prime)^2}{(2r)^2} = 5 \times 10^6.
\end{equation}

The resulting $g$-values are listed in Tab. \ref{tab:Enhancementfactor}. \\

\newpage
\section*{Author Contribution}
L.P. and G.H contributed equally to the work. D.J. helped with the optical alignment and the data analysis. G.R. developed the electrochemical etching of the Au and Ag tips. R.H. conceived the experiment. L.P., G.H., and R.H. wrote the paper.

\section*{Data availability}
The data that support the findings of this study are available from the corresponding author upon reasonable request.

\begin{acknowledgments}
This work was supported by the Deutsche Forschungsgemeinschaft (DFG) through the coordinated program TRR80 (Projekt-ID 107745057) and project HA2071/12-1. G. H. would like to thank the Alexander von Humboldt Foundation for a research fellowship. We thank S. Gepr\"ags and M. M\"uller for the fabrication of thin Au and Ag films and R. Gross for his continuous interest in the work. Finally, we gratefully acknowledge the technical support by attocube systems over the entire period during which the setup has been assembled.
\end{acknowledgments}

\bibliography{refs}

\end{document}